\documentclass[a4paper,11pt]{article}
\pdfoutput=1

\usepackage{jinstpub} 
\usepackage{graphicx}
\usepackage{microtype}
\usepackage{amssymb}
\usepackage{amsmath}
\usepackage{float}

\usepackage{lineno}

\begin{document}


\title{Use of poly(ethylene naphthalate) as a self-vetoing structural material}



\author[f]{Y. Efremenko}
\author[e]{L. Fajt}
\author[c]{M. Febbraro}
\author[a]{F. Fischer}
\author[a,b]{C. Hayward}
\author[e]{R. Hod\'ak}
\author[a]{T. Kraetzschmar}
\author[a,1]{B. Majorovits,\note{Corresponding author.}}
\author[b,2]{D. Muenstermann,\note{Corresponding author.}}
\author[a]{E. \"{O}z}
\author[g]{R. Pjatkan}
\author[d]{M. Pohl}
\author[c]{D. Radford}
\author[d]{R. Rouhana}
\author[a]{E. Sala}
\author[a]{O. Schulz}
\author[e]{I. \v{S}tekl}
\author[d]{M. Stommel}

\affiliation[a]{MPI for Physics,\\Munich, Germany}
\affiliation[b]{Lancaster University,\\Lancaster, UK}
\affiliation[c]{Oak Ridge National Laboratory, Oak Ridge,\\Oak Ridge, TN, USA}
\affiliation[d]{ TU Dortmund,\\ Dortmund, Germany}
\affiliation[e]{Institute of Experimental and Applied Physics,\\ Czech Technical University in Prague, CZ-11000 Prague, Czech Republic}
\affiliation[f]{University of Tennessee,\\ Knoxville, TN, USA}
\affiliation[g]{Nuvia a.s.,\\ T\v{r}eb\'i\v{c}, Czech Republic}

\emailAdd{bela@mpp.mpg.de,daniel.muenstermann@cern.ch}

\abstract{The discovery of scintillation in the blue regime from poly(ethylene naphthalate) (PEN), a commonly used high-performance industrial polyester plastic, has sparked the interest of the physics community as a new type of plastic scintillator material. This observation in addition to its good mechanical and radiopurity properties makes PEN an attractive candidate as an active structure scintillator for low-background physics experiments. This paper reports on investigations of its potential in terms of production tests of custom made tiles and various scintillation light output measurements. These investigations substantiate the high potential of usage of PEN in low-background experiments.}
\maketitle


\section{Introduction}
\label{S:1}

	Plastic scintillator materials with special properties such as high transparency, high structural integrity, high radiopurity have are of great interest for particle physics experiments. Such a scintillator candidate, namely,  poly(ethylene 2,6-naphthalate) or PEN, $[C_{14}H_{10}O_{4}]_n$, has been suggested as an alternative to commercial scintillators with proprietary formulas \cite{nakamura}. 
	
	PEN is a transparent polymer which has been reported to scintillate in the visible region \cite{nakamura,majorovits} without the addition of wavelength shifting dopants. The measurements of the mechanical properties of PEN both at room and cryogenic temperatures \cite{yano} have shown that both the tensile strength and the tensile modulus of PEN is simlar to that of copper, even at cryogenic temperatures; thus, it could be used as a structural material to support detectors operating in cryogenic liquids such as liquid argon (LAr) or liquid nitrogen (LN$_2$). 
	
		In this work, we investigate the properties of PEN relevant to low-background rare-event physics experiments such as Germanium Detector Array, GERDA \cite{gerda} and the Large Enriched Germanium Experiment for Neutrinoless double beta Decay, LEGEND \cite{legend}. A further background reduction could be obtained by replacing the optically inactive structural components with transparent structural plastic scintillators, such as PEN. These structural scintillating components can serve as an active veto, which would aid in discrimination of internal radioactivity as well as external background sources. The low-background copper and the Si plate detector holder used in the GERDA experiment are such examples of inactive structural components which could be replaced with structural plastic scintillator. Specifically, we report on radiopurity measurements, first production tests of custom made transparent PEN forms, measurements of their wavelength shifting and scintillating properties, and pulse-shape discrimination using PEN.

\section{Availability and radiopurity of PEN}
\begin{table*}
\caption{\label{tab:radiopurity} The results of the radiopurity measurements obtained with a low-background HPGe detector (Uncertainties are given with 68\% confidence limits).} 
\vspace{0.7cm}
\begin{tabular}{l|cccccc}
    & TN-8065S & TN-8050SC & Teonex Q51 \cite{capacitor} & CUORE link \cite{cuore_background} & Copper \cite{MajoranaCopper} & NE118 \cite{NE118}\\
    & \multicolumn{2}{c}{pellets} & foil & laminate & electroformed & scintillator\\
    & \multicolumn{6}{c}{[mBq/kg]}\\

\hline
Ra-228	&	$<$ 0.15 	 &  $<$ 0.15&\\
Th-228	&	(0.23 $\pm$ 0.05) & $<$ 0.13 & $<$ 1.4& &\\
Th-232 &           &     &     & $<$ 1.0& $<$ 12$\times 10^{-6}$ & $<$ 8.1\\
\hline
Ra-226		&(0.25 $\pm$ 0.05) & $<$ 0.11 & $<$ 2.0& &\\
Th-234	&	$<$ 11  & $<$ 15& &\\
Pa-234m&		$<$ 3.4 & $<$3.0& &\\
U-238  &        &     &        & $<$ 1.3& $<$ 7$\times 10^{-6}$ & $<$ 24\\
\hline
U-235	&	$<$ 0.066 & $<$ 0.054& &\\
\hline
K-40:		&1600 $\pm$ 400 & 1000 $\pm$ 400 & $<$ 3.6& & & $<$ 4.0  \\
Cs-137	&	$<$ 0.057 & $<$ 0.064& &\\
\end{tabular}
\end{table*}
PEN is a high-performance plastic used in many industrial applications (e.g. food packing, fibers, electronics parts, reinforced tires etc.). The polymer exhibits high chemical and hydrolysis resistance, high tensile strength, high melting point and low thermal shrinkage. In this study we investigated two types of commercial PEN acquired from Tejin-DuPont: TN-8050 SC and TN-8065 S under the brand name Teonex. These materials are provided in pellet or granulate form. Radiopurity screening was performed at the Gran Sasso National Laboratory (LNGS). The results of these measurements together with earlier screening results of Teonex Q51 film \cite{capacitor} and a laminate produced on PEN film \cite{cuore_background} are shown in Tab. \ref{tab:radiopurity}. The earlier results are from two experiments where PEN has been chosen for its radiopurity, where it was used as low-background signal links for the CUORE experiment \cite{cuore_pen}, and for the production of HV capacitors for the GERDA experiment \cite{capacitor}.

According to the LNGS results, the pellet samples have a much higher $^{40}$K contamination compared to the Teonex Q51 foil, which is suspected to be due to surface contamination of the pellets; therefore, a new screening campaign is ongoing for a new batch of TN-8065S. Apart from the increased amount of $^{40}$K, which is not a limiting factor for neutrinoless double beta decay experiments, the overall results show that TN-8065S and TN-8050SC pellets have similar radiopurity levels as PEN foil and laminate used in low-background experiments. Thus, we can conclude that PEN pellets can be used to produce high-radiopurity materials. This can potentially be improved by systematic material screening of raw ingredients prior to the synthesis process, possibly allowing for PEN to be used in areas where even lower thresholds are required.

\section{Molding of PEN}
\label{S:2}

\begin{figure}[h]
\centering\includegraphics[width=\linewidth]{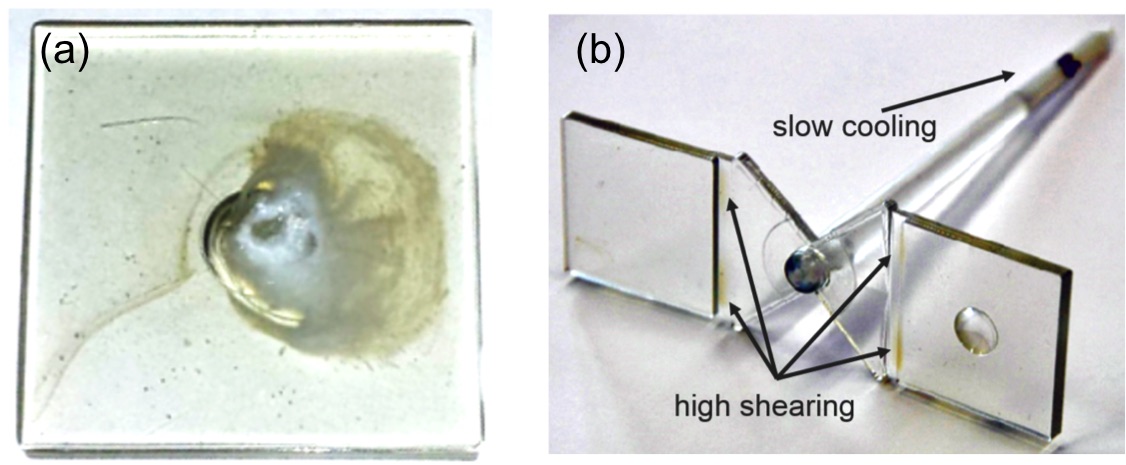}
  \caption{(a) PEN tile produced with the hot runner system. Strong opacities and other optical defects can be seen in the area of the gate. (b) PEN tiles produced with the cold runner system. A slight opacity is also visable at the boundary where film gate meets the tiles, and the crystallinity of the sprue is evident. }
  \label{fig:pentiles}
\end{figure}

 Scintillator-grade PEN tiles were produced using injection molding. Optimal manufacturing factors were identified in order to prevent crystallization and other optical defects, which reduce transparency and thus the amount of collected scintillation light. In the injection molding process, molten polymer is injected into a temperature-controlled mold under moderate to high pressure. These molds consist of an empty cavity which defines the desired geometry of the part and a channel called the runner which guides the molten polymer into the cavity. The parameters used for molding are 30$^\circ$C to 80 $^\circ$C for mold temperature, 280 $^\circ$C to 310$^\circ$C melt temperature, and an injection speed of 10 cm$^3$/s. There are two variations of the runner system, a hot runner and a cold runner. Both variants were tested for the injection molding tool used here. A hot runner guides the polymer from the machine to the mold cavity. At the end of the injection process only the tile is ejected and the hot polymer remains in the hot runner ready for the second cycle. In Fig. \ref{fig:pentiles} (a) the tile with the hot runner process can be seen without additional features, as the hot runner injects directly the polymer at the middle of the cavity. In the area of the gate, i.e. at the tip of the hot runner nozzle, strong opacities and other optical defects can be seen.
Figure \ref{fig:pentiles} (b) shows an injection molded part with a cold runner system. Here, a film gate with a smaller cross-section directly in front of the cavity is used where two tiles are produced simultaneously. This type of gate is used to balance the cavity filling. A slight opacity can be seen in this area, which is due to the increased shear. At the tip of the sprue (the channel through which the plastic is poured into the mold), the plastic cools more slowly due to the hot machine nozzle, which leads to a slower cooling rate at this point and thus increased crystallinity, hence opacity. 
\par
\begin{figure}
\centering\includegraphics[width=1\linewidth]{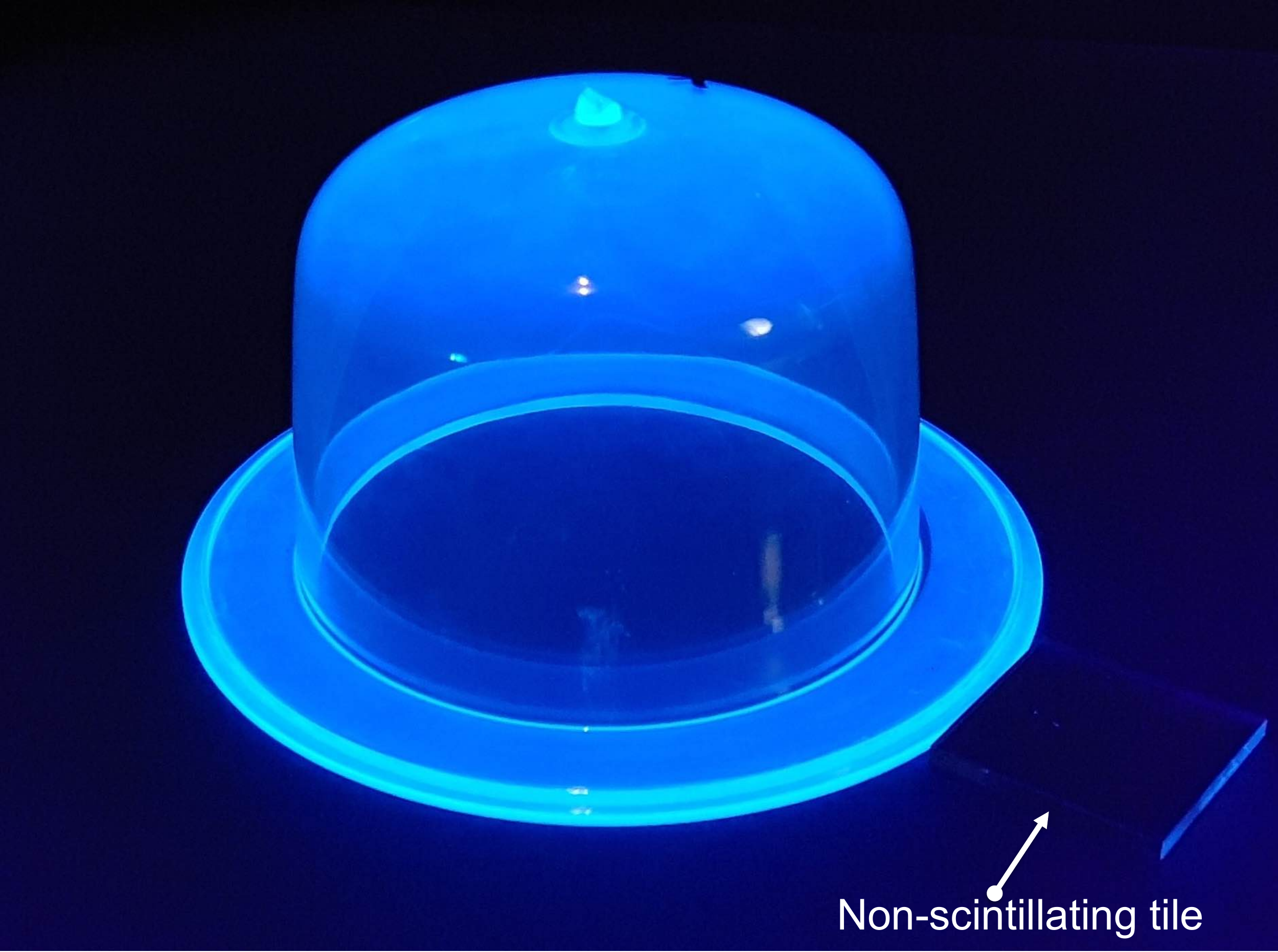}
\caption{A custom made PEN dome of height 65 mm, inner diameter 86 mm, and thickness 3 mm and a non-scintillating plastic tile (PMMA), illuminated with a UV flashlight.}
\label{fig:pencontainer}
\end{figure}

 
With these preliminary tests, an optimized geometry for the mold was developed resulting in standard-sized tiles with 30 x 30 x 3 mm$^3$ geometry. Then, a linear, fractional-factorial design of experiments was used in order to investigate the following factors for optimization: melt temperature, mold temperature, injection speed, packing pressure, packing time and cooling time. Starting parameters were determined from the manufacturer's guidelines and the previous preliminary tests. The most influential factors on the transparency were found to be the melt temperature and for the given geometry, the two-way interaction of the melt temperature and the injection speed. The highest transparency was achieved with a comparatively low melt temperature of 280 $^\circ$C and a comparatively low injection speed of 10 cm$^3$/s. These results are consistent with previous conclusions \cite{moldparameters} that the cooling rate of the melt as well as the shear stress of the melt during filling is primarily responsible for the formation of crystalline structures. These findings have also been used to produce larger shapes applicable in experiments with high-purity germanium (HPGe) detectors as an ongoing work, Fig. \ref{fig:pencontainer} shows such a form. The dome-shaped PEN of height 65 mm, inner diameter 86 mm, and thickness 3 mm is illuminated under a UV flashlight and a non-scintillating plastic tile (Methacrylate or PMMA), $[C_{5}O_{2}H_{8}]_{n}$, is placed next to it as a reference. The blue photo-luminescence from the PEN is clearly visible.

The mechanical properties of self-molded PEN tiles at room and cryogenic temperatures were measured using the three-point bending flexural test according to DIN EN ISO 178:2013-09 \cite{ISO178}. Here, the Young's modulus $E$ and yield strength $\sigma_\text{el}$ of the material are evaluated. For this measurement, the tiles were cut to $15\times 30\times 3$ mm$^3$ to fit the defined standard. The measured Young's modulus, as well as the yield strength, reach nearly double the room temperature values when submerged in liquid nitrogen. The results can be seen in Table \ref{tab:bending}. 

\begin{table*}
\centering
\caption{\label{tab:bending} Results of the three-point bending flexural test with PEN samples ($15 \times 30 \times 3$ mm$^3$) at room temperature and submerged in liquid nitrogen.}
\vspace{0.7cm}
\begin{tabular}{l|ccc}
    Temperature & PEN at $296$ K & PEN at  77  K & Copper at 296 K \cite{memsnet}\\
    \hline
    $\sigma_\text{el}$ [MPa] & $108.6 \pm 2.6$ & $209.4 \pm 2.8$ & 100\\
    $E$-Modulus [GPa] & $1.855 \pm 0.011$ & $3.708.1 \pm 0.084$ & 128 
\end{tabular}
\end{table*} 

\section{Emission spectrum}
\begin{figure}
\centering\includegraphics[width=1\linewidth]{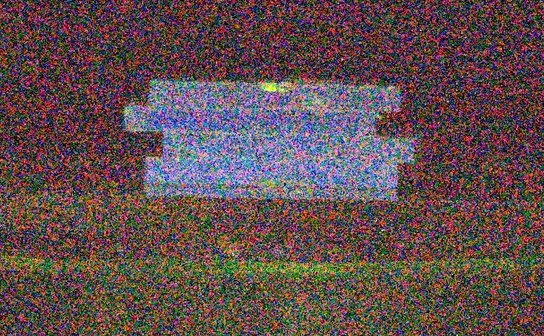}
\caption{PEN tiles photographed in the dark with a $^{137}$Cs source placed underneath the stack of tiles.}
\label{fig:scintillation}
\end{figure}
Figure \ref{fig:scintillation} shows a multi-exposure photograph of standard-sized PEN tiles obtained in a dark room under the exposure of a $^{137}$Cs source with an activity of 740 MBq. The photograph clearly shows the standard blue scintillation light.
\par
\begin{figure}
\centering\includegraphics[width=1\linewidth]{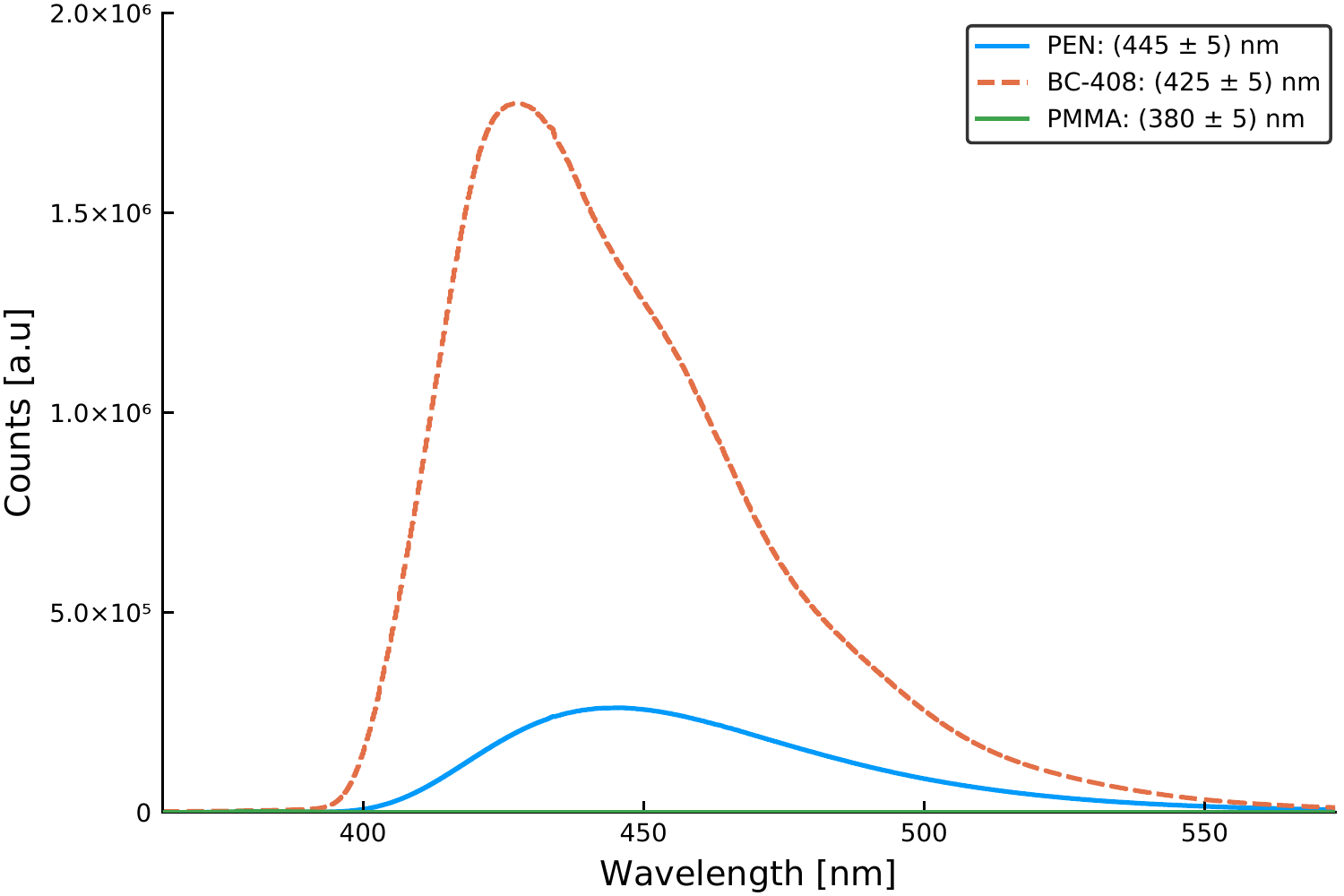}
\caption{Emission spectra of PEN, BC-408 and PMMA tiles (30 x 30 x 3 mm$^3$) with identical geometries obtained by 382 nm UV light excitation. The wavelength of the peak emission is shown in the figure's legend.}
\label{fig:emissSpec}
\end{figure}

 In order to measure the emission spectrum of PEN, a standard-size tile is excited in a dark box with a 382 nm light (with approximately $\pm$ 2 nm bandwidth). This light is generated by passing the continuous light output of a deuterium/halogen lamp (Spectral Products ASBN-D1-W) through a monochromator (Spectral Products CM110). The emitted light is then analysed using an Andor Shamrock 193i spectrograph. The Stokes shifted emission spectra of PEN, BC-408\footnote{https://www.crystals.saint-gobain.com/products/bc-408-bc-412-bc-416} and PMMA with identical geometries are shown in Fig. \ref{fig:emissSpec}. The wavelength of the emission peak of the PEN tile, which is obtained by fitting a second order polynomial to the peak region, is 445 $\pm$ 5 nm and matches well to the wavelength of the peak quantum efficiency of most commercial photomultiplier tubes (PMTs) and silicon photomultipliers (SiPMs). Note that in the literature the reported peak emission location changes slightly. This is partly due to the short absorption length of PEN. We have made preliminary measurements of the absorption length of PEN on various samples using a spectrophotometer (Thermoscientific Evolution 220). These measurements were checked for 453 nm using an independent laser absorption method. The attenuation length varies from 3 to 5 cm at 453 nm and also varies about 1 cm near the peak region (425 nm to 475 nm) with stronger absorption for shorter wavelengths. In the spectrum measurements using the monochrometer the excited light travels less than one cm before reaching the spectrograph slit; with a resulting shift in the observed peak around 3 nm.

%
%
%

\section{Vacuum-UV (VUV) response for 126.8 nm LAr scintillation light detection}
 \begin{figure}[]
\centering\includegraphics[width=1\linewidth]{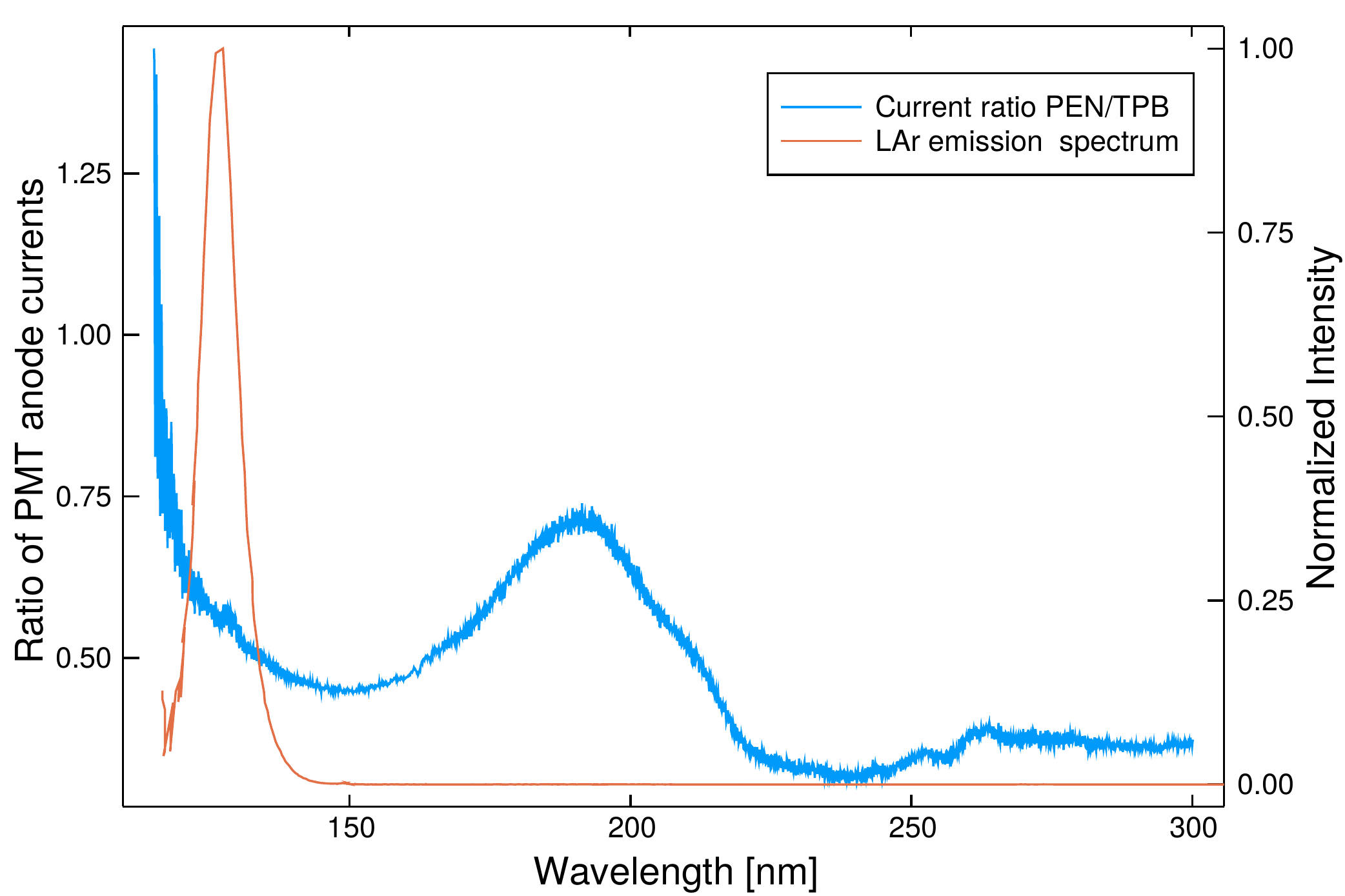}
\caption{Left axis: ratio of PMT anode currents (blue line) of a PEN tile and a TPB-coated acrylic tile excited in the range of $116$ to $300$ nm with a tunable monochrometer. The geometry of the PEN tile used corresponds to the standard format of $30\times 30\times 3$ mm$^3$ while the acrylic comparison tile measures $30\times 30\times 2$ mm$^3$ and is coated with a $200$ $\mu$m layer of TPB. Right axis: VUV/UV emission spectrum of liquid argon (85 K, orange line) with a peak emission at $126.8$ \cite{Heindl} nm.}
\label{fig:VUV}
\end{figure}
The observed Stokes shift in PEN when exposed to 382 nm light prompted further investigation down to lower wavelengths applicable to LAr scintillation detection, as liquid argon has been used as a veto and cooling in previous low background experiments. LAr scintillates in VUV with an emission peak at 126.8 nm \cite{Heindl} and thus requires either wavelength shifters which emit near blue regime or VUV-sensitive photosensors. The light yield from VUV excitation is investigated by comparing it to that of a commonly used dye, 1,1,4,4-tetraphenyl-1,3-butadiene (TPB). 200 $\mu$g/cm$^{2}$ TPB is evaporated on a 1.5 mm thick acrylic piece to produce a standard wavelength shifting TPB tile. The 3 mm thick PEN tile and the TPB coated acrylic were illuminated in a dark box with the light from a McPherson 302 VUV monochromator fitted with a Hamamatsu L1835 deuterium lamp. The wavelength of the excitation light is scanned between 116 nm and 300 nm where the wavelength uncertainty of the measurement system is less than 1 nm, and the scintillation light is collected with a PMT (Hamamatsu model R580-15 SEL ASSY). Note that the borosilicate glass window of the PMT has a cutoff at $\sim$ 300 nm therefore no excitation light enters the PMT. The PMT anode current from the PEN scintillation light normalised by that from TPB coated acrylic tile is shown in Fig. \ref{fig:VUV} along with the LAr emission spectrum. The statistical uncertainties are negligible and systematical uncertainties are estimated to be no less than 10\%. This measurement shows that while the light shifting efficiency of PEN is not as good as that of TPB, PEN can in principle be used to detect LAr scintillation light.  
 

\section{Light output comparison with mono-energetic electrons} 

\begin{figure}
\centering\includegraphics[width=1\linewidth]{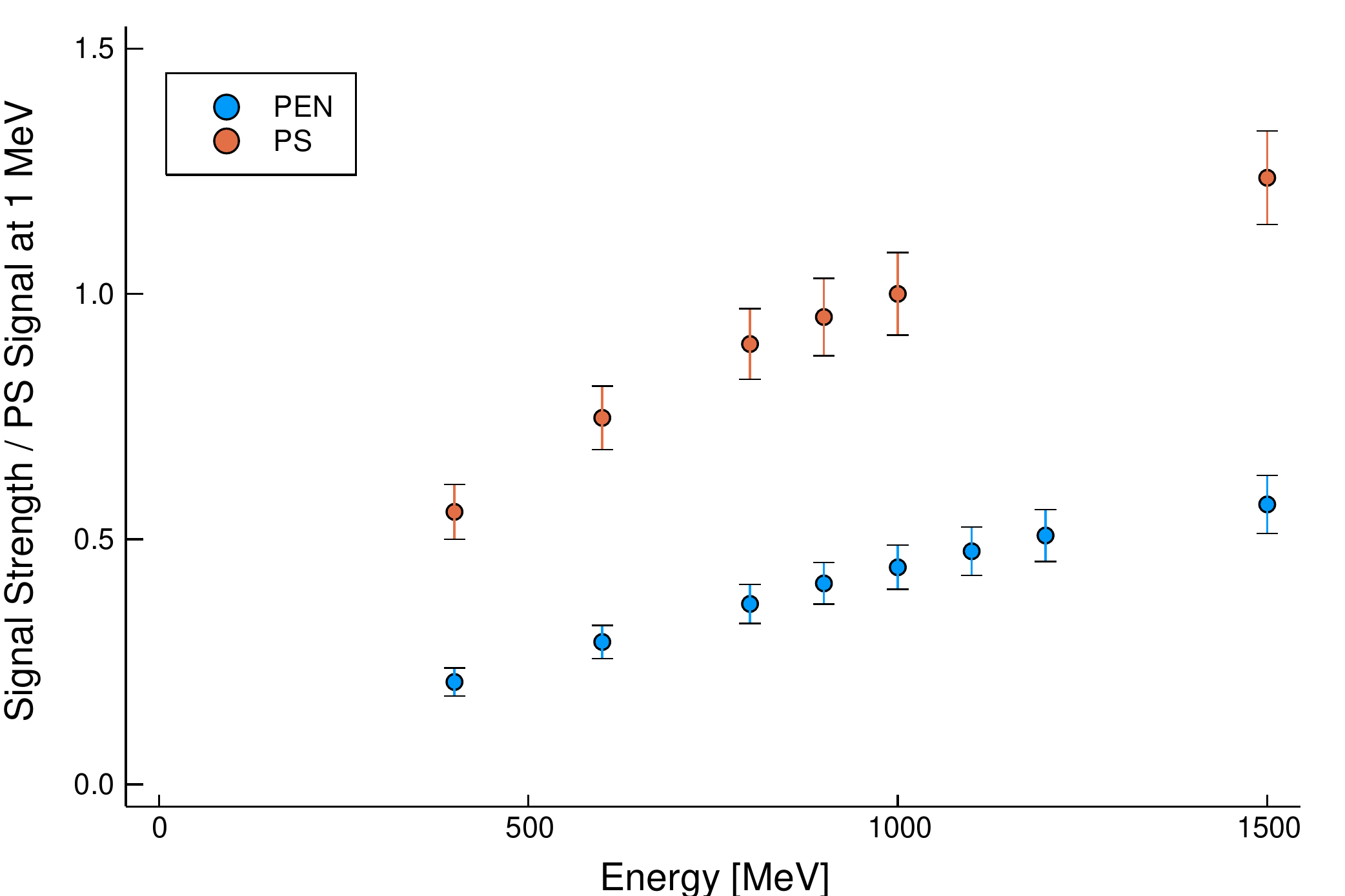}
\caption{The standard PS and PEN tiles are excited with electrons from a $^{90}$Sr source which, in combination with an electromagnet, produces narrow energy spread electrons (FWHM = 1.0 $\pm$ 0.2 \% at 1 MeV). The light output of the the two tiles of identical geometry is compared as a function of the electron energy.}
\label{fig:PSvsPEN}
\end{figure}

The scintillation light output of a standard-sized tile is compared against a commonly used plastic scintillator, polystyrene (PS)\footnote{PS is doped with the primary, \textit{para}-terphenyl (pTP), and the secondary, 1,4-bis(5-phenyloxazol-2-yl) benzene (POPOP), fluors and has the designation, SP32.} of identical geometry, which was provided by NUVIA Ltd. \footnote{https://nuvia.cz/en}. A $^{90}$Sr source in combination with an electromagnet, which acts as a velocity selector, produces tunable (0.4 to 1.5 MeV) electrons with a narrow energy spread (FWHM = 1.0 $\pm$ 0.2 \% at 1 MeV) \cite{electronspec} which is then used to excite the samples in a dark box. In order to collect the scintillation light efficiently and consistently, the samples were wrapped with a 10 $\mu$m reflective Mylar foil and then using a 1 mm thick BC-634A optical interface pad connected to a 1 inch diameter PMT (Hamamatsu model R1924A). A comparison of relative signal strengths to the PS signal at 1 MeV for various energies (see Fig. \ref{fig:PSvsPEN}) shows that a standard PEN tile emits about 2.5 times less light than the PS tile of the same geometry. Note that the non-linearities in the data set are due to lower collision cross section at higher energies. They occur at different energies due to differences in the density and effective-Z of the material.

\section{Pulse-shape discrimination}

\begin{figure}[H]\centering\includegraphics[width=1\linewidth]{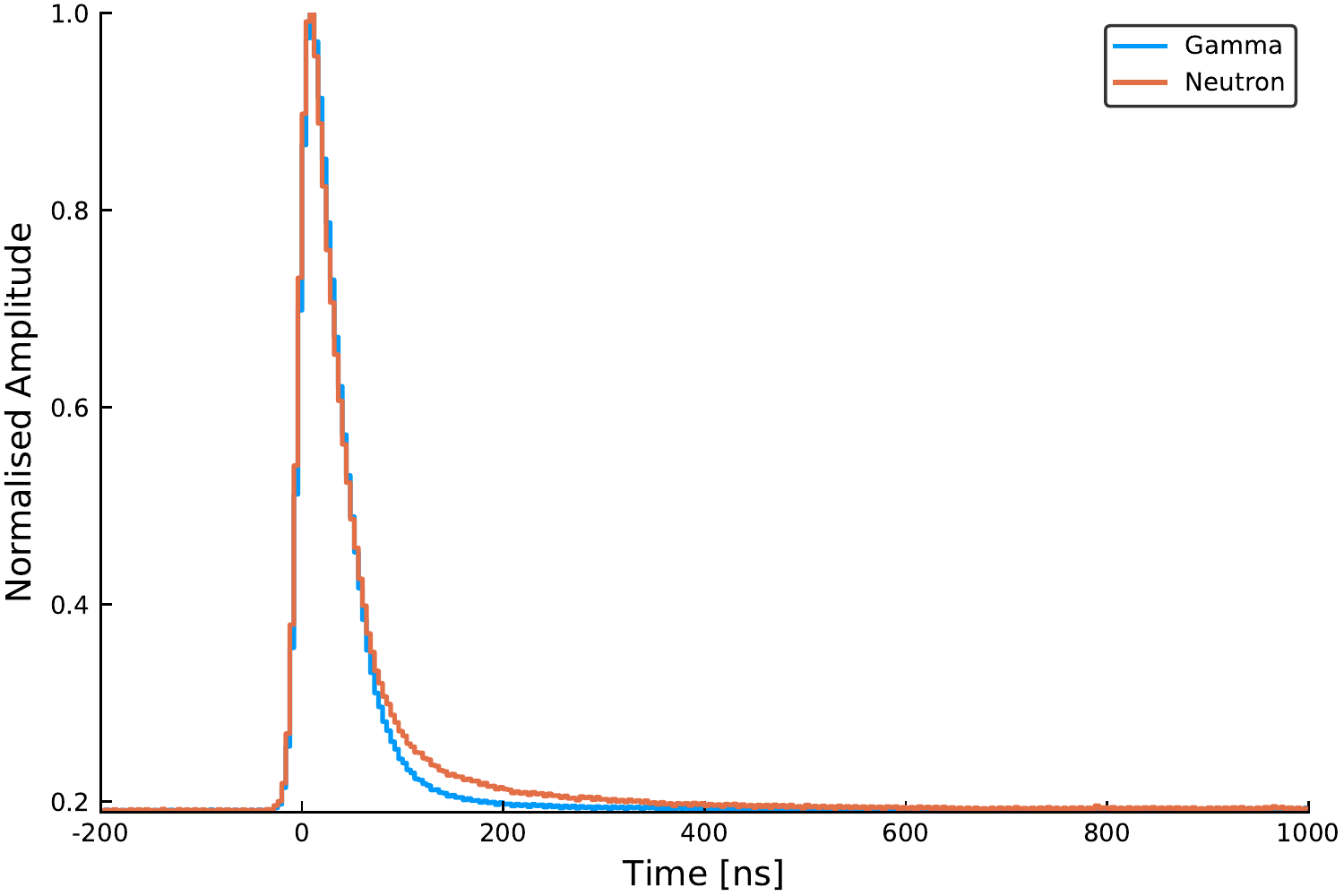}
\caption{Averaged scintillation pulse shapes of PEN for neutron (red) and gamma (blue) gated events.}
\label{fig:traces}
\end{figure}

\begin{figure}
\centering\includegraphics[width=1\linewidth]{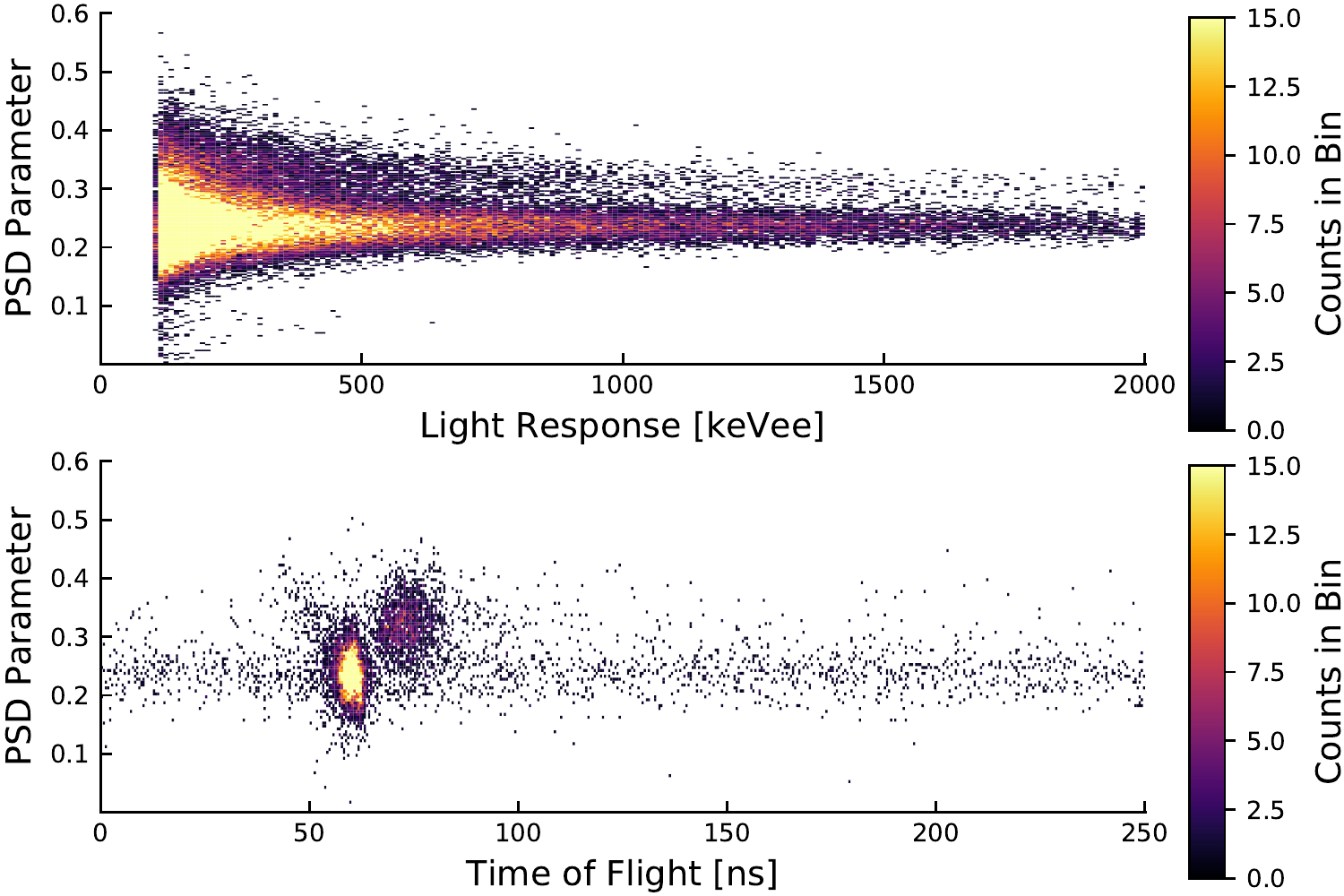}
\caption{Pulse-shape discrimination spectrum from a $^{252}$Cf fission chamber measurement from a PEN tile. For the top plot, upper and lower bands correspond to neutron interactions (recoil protons) and gamma-ray interactions (recoil electrons), respectively. The upper band is time delayed with respects to the lower bands which is indicative of neutron interactions ($v \ll c$) and gamma-ray interactions ($v = c$). The lower plot shows the pulse-shape discrimination charge ratio versus time-of-flight spectrum from a $^{252}$Cf fission chamber measurement.}
\label{fig:psd}
\end{figure}
 
\begin{figure}[h]
\centering\includegraphics[width=1\linewidth]{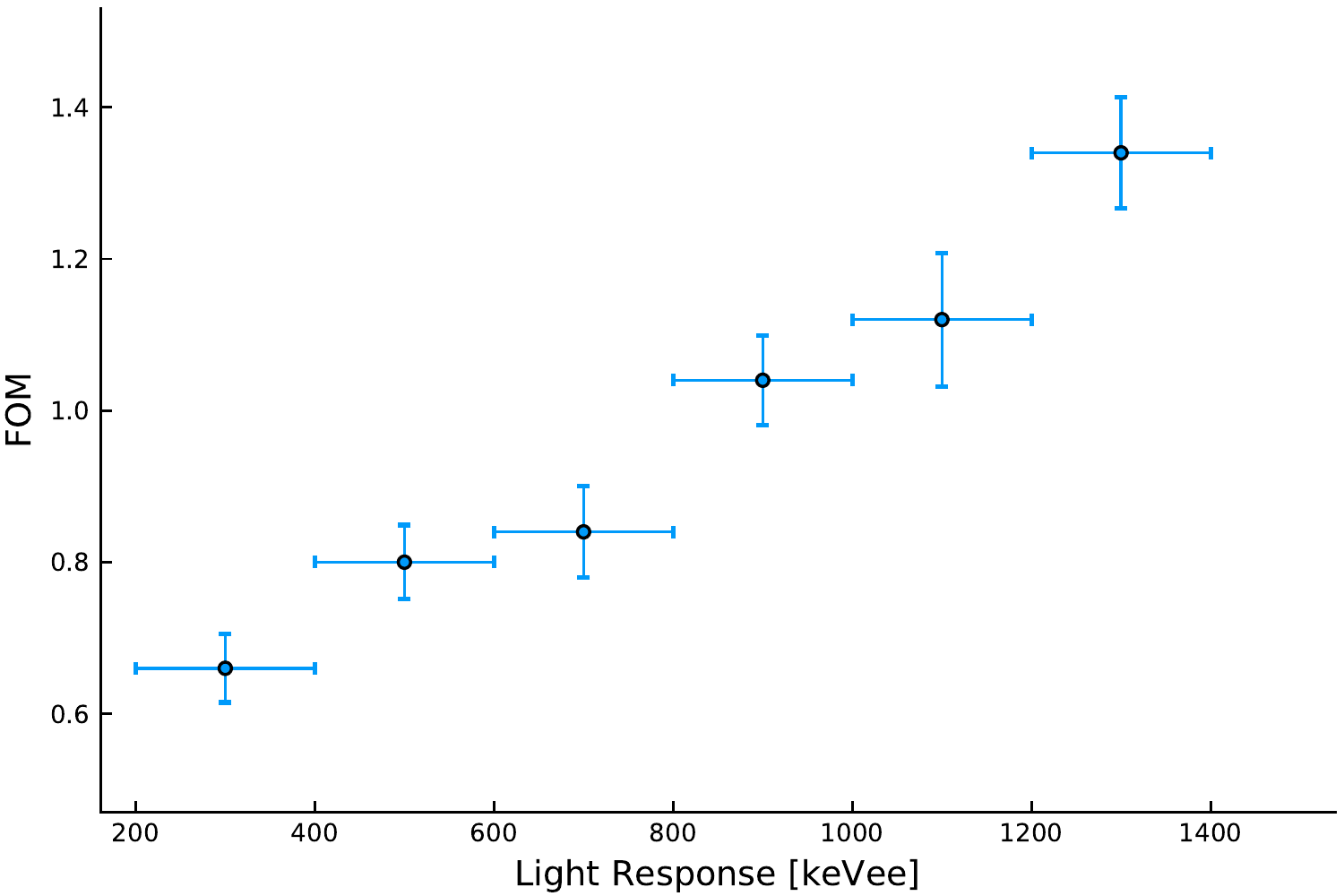}
\caption{FOM vs light response for the pulse-shape discrimination measurement of gammas and neutrons.}
\label{fig:fom}
\end{figure}

Depending on the density of the ionisation deposited in the material, the intensity of the emitted light can have different time dependency, i.e. different pulse shapes. For example, a heavy particle such as a neutron will cause a higher ionisation or excitation density which will reduce the scintillation efficiency and cause a longer/slower decay time, and hence pulse shape. Two such pulse shapes separated by gating are shown in Fig. \ref{fig:traces}. Pulse-shape discrimination (PSD) methods take advantage of this feature to distinguish the type of excitation in low-background experiments, and provide information to remove these events. A $^{252}$Cf fission chamber was used at Oak Ridge National Laboratory in order to test a PSD method based on charge integration on a standard PEN tile. The $^{252}$Cf fission chamber also provides time-tagged gamma rays and fast neutrons which can be used to evaluate the time of flight and hence the quality of the PSD method. The 30 x 30 x 3 mm PEN tile is wrapped in PTFE tape and optically coupled to a PMT (Hamamatsu model R6231-100-01 ASSY) using BC-630 optical grease. The PMT signal is digitised at 250 MHz and at 14-bit using a CAEN V1725 waveform digitiser. The PSD parameter, i.e. the charge integration ratio (or tail-to-total) is extracted from the signal where the total integral is over 850 ns and the tail integral is over 750 ns starting from 50 ns from the leading edge of the pulse. The PSD parameter is plotted against the light response in Fig. \ref{fig:psd}. A clear two band structure that can be used to discriminate neutrons and gammas is visible. This is consistent with the time of flight measurement shown in Fig. \ref{fig:psd}. In order to quantify the discrimination quality, we further calculate the figure-of-merit (FOM) \cite{fom} which is defined as: 
\begin{equation}
FOM=\frac{ \mid X_\gamma - X_n\mid}{W_\gamma+W_n}
\end{equation}
where X represents the peak position of the PSD parameter of the gammas or neutrons and W represents their respective FWHM. The result as a function of light response is shown in Fig. \ref{fig:fom}. FOM is clearly better for higher light response. Similar pulse discrimination calculations has also been reported previously in Ref. \cite{PSD_PEN_paper} where 125 $\mu$m PEN films were exposed to neutrons and gammas up to 10 MeV from $^{239}$PuBe and $^{238}$PuBe source and the FOM was measured as 0.69 $\pm$ 
0.02 for energies up to 14 MeV. For applications in low-background experiments the PSD capability of PEN is a great advantage as it allows for achieving additional information on the type of interaction and hence helps identifying background sources.

\section{Conclusions \& Outlook}
In summary, the following properties of PEN have been identified in our investigations: HPGe screening measurements have shown that
commercially available PEN pellets with radiopurities well below 1mBq/kg in $^{228}$Th and $^{226}$Ra can be used to produce structural parts.
 Molding tests indicate that scintillator-grade arbitrary shapes such as dome-shaped structures can be produced with ease. In addition, the newly measured peak emission wavelength of PEN scintillation light (445 $\pm$ 5 nm) matches well to that of the peak quantum efficiency of most PMTs and SiPMs. The comparison of light shifting efficiency of PEN for LAr scintillation light with that of TPB show that PEN can in principle be used to shift LAr scintillation light to the wavelength readily accessible to PMTs and SiPMs. According to the results of the study of the light output of PEN using mono-energetic electrons, PEN has a linear response as a function of energy. The sample used in the investigation emitted about 2.5 times less light than a PS scintillator. There is also a discrepancy between these results and previous work \cite{nakamura}. This is believed to be due to the relatively low attenuation length of the self molded PEN sample, of order 5 cm, and further efforts to quantify the light yield without the effects of attenuation are ongoing. Lastly, the results of the pulse-shape discrimination studies of PEN show that PEN scintillation pulses can be used to discriminate neutrons and gammas with a good efficiency for the energy range of 300 to 1300 keVee, where ee stands for electron equivalent. Therefore, the combination of all of the above mentioned properties make PEN a very promising candidate as a structural material in low-background experiments, especially where optically active materials are desirable. Presently, 
efforts are underway to self synthesize PEN from carefully selected and purified ingredients. Also, measurements are being performed to further quantify the light yield, attenuation length, and wavelength shifting behavior. Also a first setup to operate a high-purity germanium detector surrounded by a custom made PEN enclosure directly in LAr is being prepared.

\acknowledgments
This work was supported by the Ministry of Industry and Trade of the Czech Republic under the Contract Number FV30231. Research sponsored by the Laboratory Directed Research and Development Program of Oak Ridge National Laboratory, managed by UT-Battelle, LLC, for the U.S. Department of Energy. This material is based upon work supported by the U.S. Department of Energy, Office of Science, Office of Nuclear Physics, under Award Number DE-AC05-00OR22725. Erdem \"{O}z and Elena Sala are supported by the Deutsche Forschungsgemeinschaft (SFB1258). Connor Hayward is supported by the Excellence Cluster Universe.








\end{document}